\documentclass[letterpaper, 10 pt, conference]{ieeeconf} 
\usepackage{subcaption}
\usepackage{graphicx}
\usepackage{changepage}
\usepackage{booktabs}
\usepackage{afterpage}
\usepackage{balance}
\usepackage{float}
\usepackage{amssymb} 
\usepackage{multirow}
\usepackage{diagbox}

\IEEEoverridecommandlockouts 

\usepackage{amsmath} 

\title{\LARGE \bf
HIPPO-MAT: Decentralized Task Allocation Using GraphSAGE and Multi-Agent Deep Reinforcement Learning
}

\author{Lavanya Ratnabala \textsuperscript{1,*}, Robinroy Peter \textsuperscript{2,*}, Aleksey Fedoseev \textsuperscript{1},  and Dzmitry Tsetserukou \textsuperscript{1}
\thanks{\textsuperscript{*}The authors contributed equally to this work.}
\thanks{\textsuperscript{1}The authors are with the Intelligent Space Robotics Laboratory, Skolkovo Institute of Science and Technology, Bolshoy Boulevard 30, bld. 1, 121205, Moscow, Russia.}
\thanks{\{\tt lavanya.ratnabala, aleksey.fedoseev, d.tsetserukou\}@skoltech.ru} 
\thanks{\textsuperscript{2}The author is with NeuroFleets (PVT) LTD, Jaffna, Sri Lanka.}
\thanks{\tt\ robinroy.peter@neurofleets.com}  
}

\begin{document}

\maketitle
\thispagestyle{empty}
\pagestyle{empty}


\begin{abstract}
This paper tackles decentralized continuous task allocation in heterogeneous multi-agent systems. We present a novel framework HIPPO-MAT that integrates graph neural networks (GNN) employing a GraphSAGE architecture to compute independent embeddings on each agent with an Independent Proximal Policy Optimization (IPPO) approach for multi-agent deep reinforcement learning. In our system, unmanned aerial vehicles (UAVs) and unmanned ground vehicles (UGVs) share aggregated observation data via communication channels while independently processing these inputs to generate enriched state embeddings. This design enables dynamic, cost-optimal, conflict-aware task allocation in a 3D grid environment without the need for centralized coordination. A modified A* path planner is incorporated for efficient routing and collision avoidance. Simulation experiments demonstrate scalability with up to 30 agents and preliminary real-world validation on JetBot ROS AI Robots, each running its model on a Jetson Nano and communicating through an ESP-NOW protocol using ESP32-S3, which confirms the practical viability of the approach that incorporates simultaneous localization and mapping (SLAM). Experimental results revealed that our method achieves a high 92.5\% conflict-free success rate, with only a 16.49\% performance gap compared to the centralized Hungarian method, while outperforming the heuristic decentralized baseline based on greedy approach. Additionally, the framework exhibits scalability with up to 30 agents with allocation processing of 0.32 simulation step time and robustness in responding to dynamically generated tasks.
\end{abstract}

\section{Introduction}

Multi-agent systems offer significant advantages in terms of scalability, distributed operation, and parallel task execution in modern automated logistics systems. However, efficiently assigning continuously emerging tasks to the most suitable agents remains a formidable challenge, particularly in dynamic 3D environments where simultaneous decision-making is required and conflicts can arise when multiple agents select the same task.

\begin{figure}[ht]
 \centering
 \includegraphics[width=1.0\linewidth]{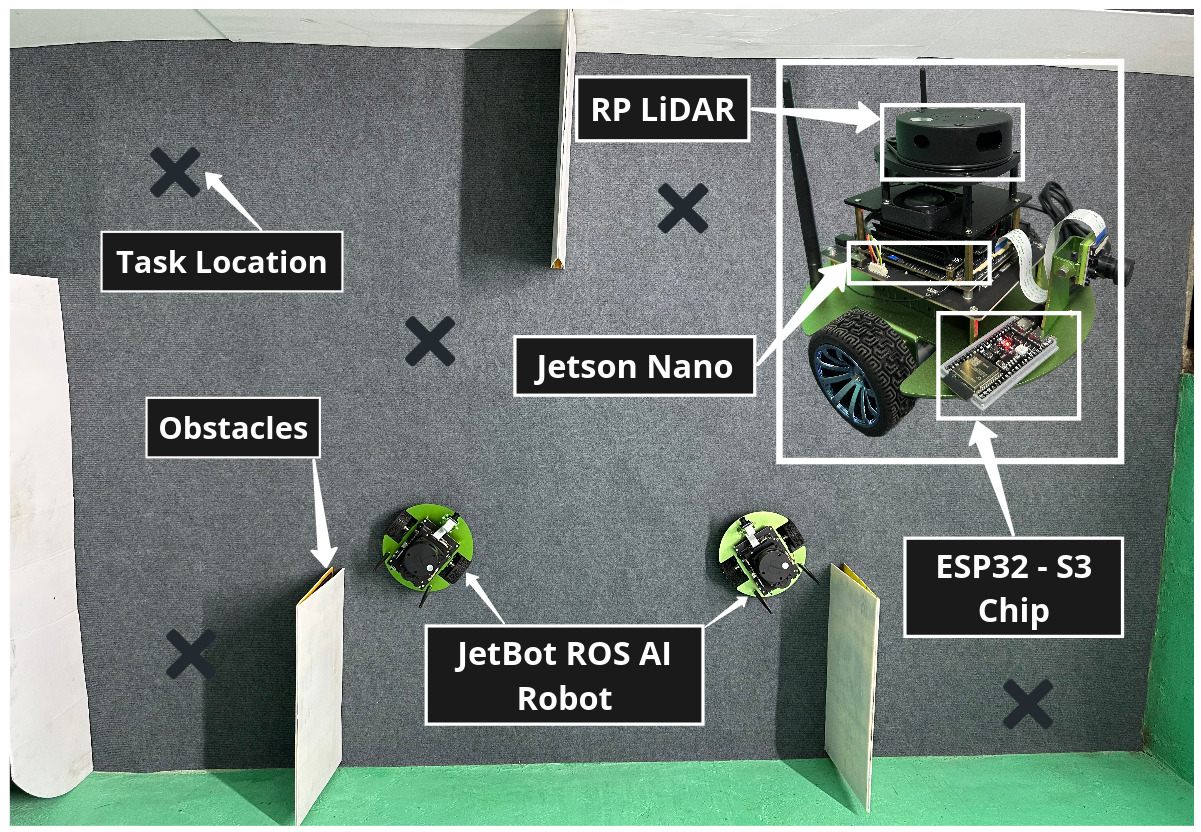}
 \caption{Jetbot ROS AI Robots equipped with Jetson Nano for individual policy loading and ESP32-S3 for inter-agent communication.}
 \label{fig:real}
\end{figure}

Traditional task allocation approaches are often divided into centralized and decentralized methods. Centralized techniques, such as the Hungarian algorithm \cite{Kuhn_1955}, can produce globally optimal assignments but require continuous high-bandwidth communication and are prone to bottlenecks and single points of failure. Decentralized methods, on the other hand, typically operate in 2D and assume sequential or one-by-one task processing. This sequential assumption, however, is impractical in real-world scenarios where tasks must be processed concurrently. Moreover, existing decentralized approaches often yield suboptimal allocations, leading to conflicts or redundant task assignments and increased overall costs. and most of them doesn't touch on conflict management. 

We present a novel framework aimed at improving real-time task allocation. Our approach enables continuous task allocation in a three-dimensional environment through concurrent, decentralized decision-making. We integrate a graph neural network (GNN) based on a GraphSAGE architecture \cite{graphsage} to compute independent state embeddings on each agent. These embeddings capture critical inter-agent and agent-to-task interactions, allowing each robot to evaluate the shared task pool concurrently and resolve conflicts by ensuring unique task assignments. By coupling these GNN-enhanced representations with an Independent Proximal Policy Optimization (IPPO), our method optimizes task allocation decisions to minimize total cost (e.g., travel distance and time) while ensuring efficient routing via a modified A* path planner.


In summary, our main contributions are: \begin{itemize}
\item Concurrent Task Allocation in 3D: We develop a framework that enables multiple agents to concurrently access and allocate continuously emerging tasks in a dynamic three-dimensional environment, overcoming the limitations of simultanious robots processing the tasks in Decentralised MATA. 
\item GNN-Enhanced Independent Policy Learning: Novel approach By integrating GraphSAGE-based independent state embeddings with an IPPO framework, our approach captures rich inter-agent interactions, enabling informed, near optimal conflict-aware task assignment decisions.  
\item Coupled Navigation and Real-World Validation: Our framework tightly couples task allocation with reservation-based A* path planner to optimize routing and collision avoidance, and we validate its performance through extensive simulations and preliminary real-world experiments using JetBot ROS AI Robot with SLAM-based navigation. 
\end{itemize}

\section{Related Works}

Optimization techniques, e.g., the linear sum assignment problem (LSAP) and the Hungarian algorithm, have been widely applied in multi-agent task allocation \cite{goarin2024graph}. While these methods provide optimal solutions, they often rely on centralized computation, rendering them unsuitable for large-scale real-time applications. Ismail et al. \cite{ismail2017decentralized} proposed a novel decentralized-based Hungarian method to solve this problem. Another decentralized version of the Hungarian method is proposed by Xia et al. \cite{xia} to solve task allocation for underwater vehicles. These methods extend the Hungarian approach to decentralized settings by ensuring task allocation remains optimal as long as agent networks remain connected. Kong et al. \cite{greedy} proposed a new optimization strategy that combines the improved particle swarm optimization and the greedy IPSO-G algorithm. Moreover, two different stochastic approaches, the Genetic Algorithm (GA) and the Ant-Colony Optimization (ACO) algorithm, were introduced by \cite{genetic} to solve the multi-robot task allocation problem. Peter et al. \cite{neurofleets} introduced a swarm intelligence-based task allocation method utilizing decentralized decision-making and subgoal-based path formation.

Market-based strategies, particularly auction-based methods, have been widely explored for decentralized task allocation. Zhong et al. \cite{zhong2018stable} proposed an extended auction-based driver-passenger matching system for ride-sharing applications. Liu et al. \cite{liu2024multi} extended auction-based methods to multi-UAV systems, incorporating Dubins path-based flight cost estimation to refine task allocations. While auction-based strategies provide high adaptability, their efficiency heavily depends on accurate bid valuation models. An alternative hybrid approach is introduced in the Harmony Drone Task Allocation (DTA) method \cite{harmonyDTA}, which integrates a consensus-based auction mechanism with a gossip-based consensus strategy. The Harmony DTA optimizes multi-drone task allocation under complex time constraints by balancing task urgency with resource availability while minimizing communication load.

The integration of DRL into MATA has led to promising advancements, particularly through MARL frameworks. Agrawal et al. \cite{agrawal2023rtaw} introduced an attention-inspired DRL method for warehouse-based task allocation that optimizes decision-making efficiency and scalability. Similarly, Dai et al. \cite{dai2025scheduling} propose a decentralized RL approach for heterogeneous multi-robot task allocation, where robots with diverse skills form coalitions to complete tasks, minimizing the overall mission completion time. They use attention mechanisms to reason about task dependencies and agent skills and introduce a constrained flash forward mechanism to improve training efficiency.

Moreover, recent studies explore Multi-Agent Proximal Policy Optimization (MAPPO) to enhance collaborative decision-making and mitigate non-stationary challenges in multi-agent environments \cite{mappo2023arxiv}. Lowe et al. \cite{lowe2017multi} use Q-learning to do cooperative and competitive tasks. However, Christian et al. show that IPPO, where each agent learns its own policy using only local observations, can achieve performance on par with or even better than state-of-the-art centralized training methods (e.g., QMIX, MAPPO) on several challenging maps from the StarCraft Multi-Agent Challenge (SMAC) \cite{ippo}.

Recent advances in graph-based learning methods have demonstrated substantial potential for improving decentralized task allocation. GNNs enable robots to model and process inter-agent relationships, allowing task allocations to be computed based on local graph structures rather than centralized control. Goarin et al. \cite{goarin2024graph} proposed DGNN-GA, a decentralized GNN-based goal assignment method that optimizes communication efficiency in multi-robot planning for a fixed number of tasks. Heterogeneous multi-agent systems pose additional complexities due to differences in robot capabilities, sensor modalities, and mobility constraints. The GATAR framework \cite{peng2024graph} introduced a graph-based task allocation framework for multi-robot target localization, specifically designed for heterogeneous robot systems. Furthermore, Blumenkamp et al. \cite{blumenkamp2022framework} develop a real-world framework for deploying decentralized GNN-based policies in multi-robot systems, facilitating seamless sim-to-real transfers.  Bettini et al. propose Heterogeneous Graph Neural Network Proximal Policy Optimization (HetGPPO), a framework for training heterogeneous MARL policies using Graph Neural Networks (GNNs) for inter-agent communication. By enabling agents to learn distinct behaviors while maintaining fully decentralized training in partially observable environments, HetGPPO highlights the effectiveness of GNNs in decentralized MARL \cite{Hetgppo}. Ratnabala et al. \cite{Ratnabala_2025} recently proposed a GNN-based DRL framework for heterogeneous robot swarms. The suggested MAGNNET approach allowed an efficient compromise between optimization methods and locally optimized greedy approaches; however, the centralized critic model suggests a challenge in further scalability of the system. 

Despite significant progress, current approaches exhibit several key limitations. Many continuous task allocation methods assume that tasks are processed sequentially, which is impractical in scenarios where multiple agents must simultaneously access, evaluate, and assign tasks. Additionally, most existing studies \cite{agrawal2023rtaw} assume that only one robot is available at a time and skip the conflict resolution part, failing to capture the complex dynamics of three-dimensional spaces essential for realistic navigation and task allocation. Moreover, task allocation is often decoupled from navigation, even though efficient systems require integrated routing and collision avoidance. Finally, while some methods address heterogeneous teams, they typically rely on centralized information or static task models, limiting scalability in dynamic, large-scale deployments. In contrast, our work proposes a decentralized framework for continuous task allocation in 3D environments by integrating a GraphSAGE-based GNN with an Independent PPO (IPPO) framework. This approach enables multiple agents to concurrently access and optimize a shared task pool, resolve conflicts, and minimize overall cost while coupling task allocation with a modified A* path planner for efficient navigation and collision avoidance.

\section{Multi-Agent Task Allocation Approach}
\subsection{Problem Formulation}
We consider a system of \(N\) agents (or robots), denoted by \(\{a_1, a_2, \dots, a_N\}\), operating in a continuous three-dimensional domain \(\Omega \subset \mathbb{R}^3\). A set of \(M\) tasks, \(\{T_1, T_2, \dots, T_M\}\), is generated over time. Each task \(T_j\) is characterized by its location \(\mathbf{l}_j \in \Omega\) and a status \(s_j(t) \in \{\text{Waiting}, \text{Assigned}\}\) at time \(t\). Once a task is allocated to an agent, it is replaced by a new task.

Each agent \(a_i\) is described by its position \(\mathbf{p}_i(t) \in \Omega\) and its operational status (e.g., \emph{idle}, \emph{accept}, \emph{assign}, or \emph{complete}). The travel cost for agent \(a_i\) to complete task \(T_j\) at time \(t\) is defined as:
\begin{equation}
 c_{ij}(t) \;=\; \frac{d_{ij}(t)}{v_i},
\end{equation}
where \(d_{ij}(t)\) is the shortest-path distance (e.g., computed via A*) from \(\mathbf{p}_i(t)\) to \(\mathbf{l}_j\), and \(v_i\) is the velocity of \(a_i\). The cost values are clipped to the interval \([0, c_{\max}]\) and then normalized to \([-1, 1]\) for improved training stability.

Rather than solving for a centralized assignment using binary variables \(x_{ij}\in\{0,1\}\), each agent selects an action from a discrete set:
\begin{equation}
 A_i \;=\; \{0,1,\dots,M\},
\end{equation}
where action “0” denotes absence of the task request, action \(j \in \{1,\dots,M\}\) denotes the request of the task \(T_j\). When multiple agents request the same task, the task is allocated to the agent with the smallest normalized cost, while the other agents incur a penalty.

\subsection{Graph Neural Network for Independent Agent Embedding}
We model the multi-agent system as a graph, \(\mathcal{G} = (\mathcal{V}, \mathcal{E})\) where each node \(v_i \in \mathcal{V}\) corresponds to an agent \(a_i\). The graph is fully connected, meaning that for every pair of distinct agents \(a_i\) and, \(a_j\) there exists an edge \((v_i, v_j) \in \mathcal{E}\). This connectivity allows each agent to aggregate information from all other agents in the system.

The raw observation for each agent \(a_i\) is represented by a vector \(\mathbf{x}_i \in \mathbb{R}^{1+2M}\). This vector consists of three elements: the agent’s status (a scalar normalized to \(-1\) when idle and \(+1\) when assigned), a set of \(M\) normalized cost values corresponding to the travel costs to each available task, and a set of \(M\) binary indicators that denote the status of each task (with \(+1\) for a waiting task and \(-1\) for an inactive task). Thus, \(\mathbf{x}_i\) captures the local state of the agent with respect to task allocation.

The edge set \(\mathcal{E}\) is defined as:
\[
\mathcal{E} = \{(v_i, v_j) \mid \forall\, i,j \in \{1,\dots,N\},\, i \neq j\},
\]
which means that each agent is directly connected to every other agent. No explicit edge features are used; rather, the connectivity enables the diffusion of information across the network.

To obtain a compact and informative representation of each agent’s state, we apply a GraphSAGE layer. The embedding for agent \(a_i\) is computed as:
\[
\mathbf{z}_i \;=\; \tanh\!\Biggl( \mathbf{W}\,\mathbf{x}_i + \sum_{j \neq i} \mathbf{W}'\,\mathbf{x}_j \Biggr),
\]
where \(\mathbf{W}\) and \(\mathbf{W}'\) are the learnable weight matrices. \(\mathbf{W}\,\mathbf{x}_i\) is the self-transformation of the agent's own features, while the summation \(\sum_{j \neq i} \mathbf{W}'\,\mathbf{x}_j\) aggregates the features from all other agents. The \(\tanh\) activation function is applied to normalize the output to the range \((-1, 1)\). The final embedding \(\mathbf{z}_i \in \mathbb{R}^{6}\) is then used as the input to the agent’s policy network.

This graph-based formulation, with its fully connected structure and GraphSAGE aggregation, enables each agent to indirectly incorporate global information about the states of all other agents. Notably, they generate independent embeddings. Consequently, even though each agent operates in a decentralized manner, it benefits from a rich contextual representation that supports effective and coordinated decision-making in dynamic environments. These final embeddings serve as input to the policy network.

\subsection{Independent PPO Training \& Decentralized Execution}
\subsubsection{Independent Training}
Each agent \(a_i\) is equipped with an individual policy \(\pi_i\) and a corresponding value network, and training is conducted using Independent Proximal Policy Optimization (IPPO). At each time step \(t\), agent \(a_i\) observes its state as:
\[
 \mathbf{o}_i(t) = \bigl[\mathbf{x}_i(t),\, \mathbf{z}_i(t)\bigr].
\]

\begin{figure*}[tp]
 \centering
 \smallskip
 \includegraphics[width=0.85\textwidth]{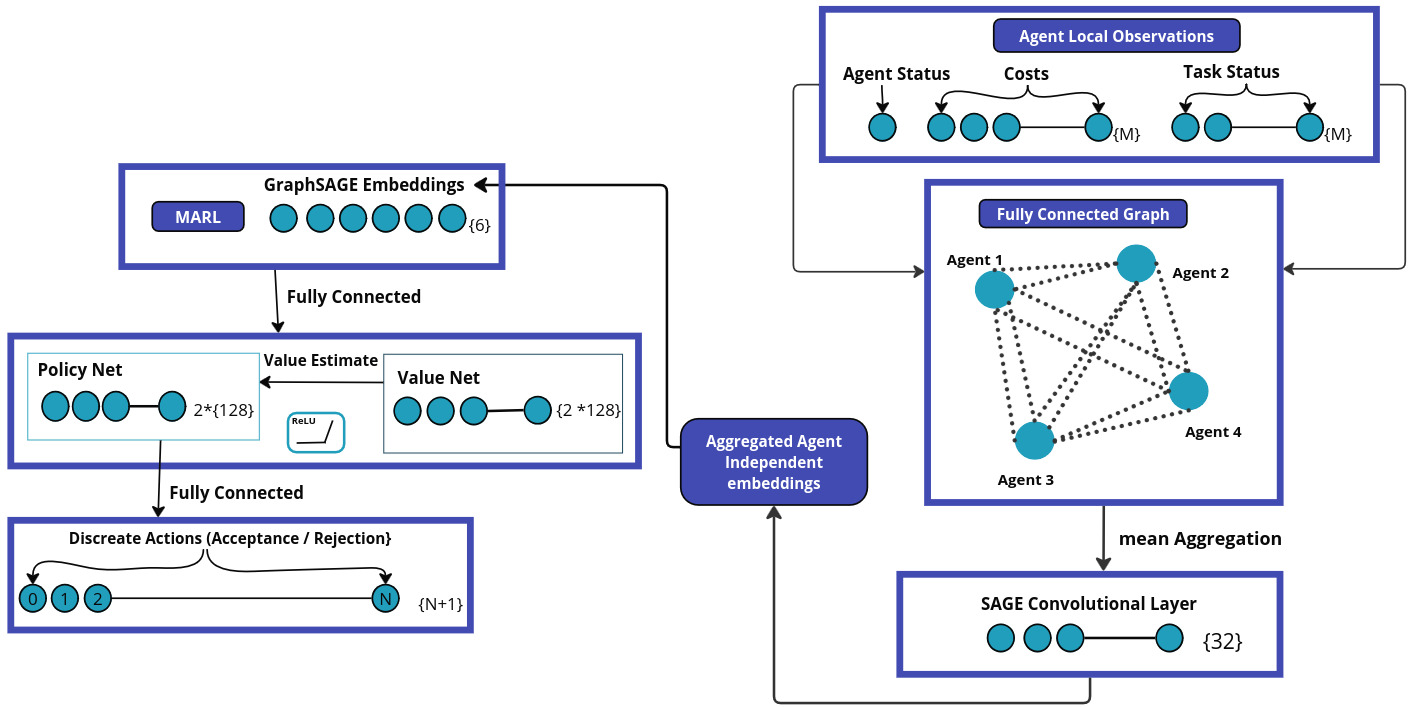}
 \caption{Multi-agent reinforcement learning and GraphSAGE architecture for task allocation.}
 \label{fig:land}
\end{figure*}

after that, it selects an action \(a_i(t) \in A_i\). Task assignments are determined by comparing the normalized costs, with the agent exhibiting the minimum cost only receiving a positive reward, while others incur a conflict penalty. The instantaneous reward for agent \(a_i\) is given by:
\begin{equation}
  r_i(t) \;=\; 
  \begin{cases}
    -c_{ij}(t), & \text{if } a_i \text{ is assigned task } T_j, \\
    -\lambda,   & \text{if multiple agents request } T_j \text{(conflict)}, \\
    -\mu,      & \text{if } a_i \text{ idle status without valid reason}, \\
    +\eta,     & \text{low-cost assignment is achieved,}
  \end{cases}
\end{equation}
where \(\lambda\), \(\mu\), and \(\eta\) are the constants regulating the penalties and bonus. A global reward signal is calculated as:

\begin{equation}
  R(t) \;=\; \sum_{i=1}^{N} r_i(t).
\end{equation}

\subsubsection{Decentralized Execution}
After training, each agent executes its learned policy \(\pi_i\) using only its local observation \(\mathbf{o}_i(t)\). No centralized coordinator is required, and decision-making is fully decentralized, enhancing scalability and real-time responsiveness.

\subsection{Model Architecture and Training Strategy}
\subsubsection{Neural Network Architecture}

In our IPPO framework, each agent’s policy and value networks receive as input the final GraphSAGE embedding \(\mathbf{z}_i \in \mathbb{R}^6\), which is computed from the agent’s raw, normalized observation. The policy network is implemented as a multi-layer perceptron (MLP) consisting of two fully connected hidden layers, each with 128 neurons and ReLU activations. The output layer applies a softmax activation to produce a probability distribution over the discrete action space \(A_i = \{0,1,\dots,M\}\). Similarly, the value network responsible for estimating the expected return \(V_i(t)\) employs an analogous architecture with two hidden layers of 128 neurons (using ReLU activations) and concludes with a linear output layer. In our independent PPO setup, each agent’s networks are updated separately based solely on its local experience.

Both networks are optimized using the Adam optimizer with a learning rate of \(1 \times 10^{-5}\). Training is conducted with a rollout fragment length of 100 steps, a batch size of 1000, and 10 SGD iterations per update. An entropy coefficient of 0.05 is applied to promote exploration, while a discount factor \(\gamma = 0.99\) and Generalized Advantage Estimation (GAE) with \(\lambda = 0.95\) are used to stabilize policy updates. The overall neural network architecture is shown in Fig.~\ref{fig:land}.

\subsubsection{Reward Shaping and Task Replacement}
The reward structure is designed to encourage optimal task allocation by directly incorporating the travel cost into the reward signal. Specifically, if an agent \(a_i\) is assigned a task \(T_j\), it receives a reward equal to \(-c_{ij}(t)\), where \(c_{ij}(t)\) is the travel cost from \(a_i\) to \(T_j\). Because lower travel costs yield fewer negative reward values, an agent that selects a closer task effectively minimizes its penalty, thereby promoting optimal task assignments.

In cases where multiple agents request the same task, only the agent with the minimum cost is receiving the cost-based reward, while the other agents incur a high penalty of \(-\lambda\). Moreover, if an agent remains idle without justification, it is penalized by \(-\mu\). A bonus \(\eta\) is awarded when an assignment is both efficient and low-cost. If already assigned agent, try to accept new while not completing the current task will get additional penalty.

Notably, once a task is allocated, it is immediately replaced via a randomized procedure (as described in Section~\ref{sec:simulation}), ensuring that agents are continuously exposed to diverse scenarios during training. This reward design, in conjunction with the independent policy updates in our IPPO framework, drives agents to learn decentralized strategies that favor a minimal travel cost for task allocation.

\subsection{Simulation Setup and Path Planning}
\label{sec:simulation}
\begin{figure}[ht]
 \centering
 \smallskip
 \includegraphics[width=1.0\linewidth]{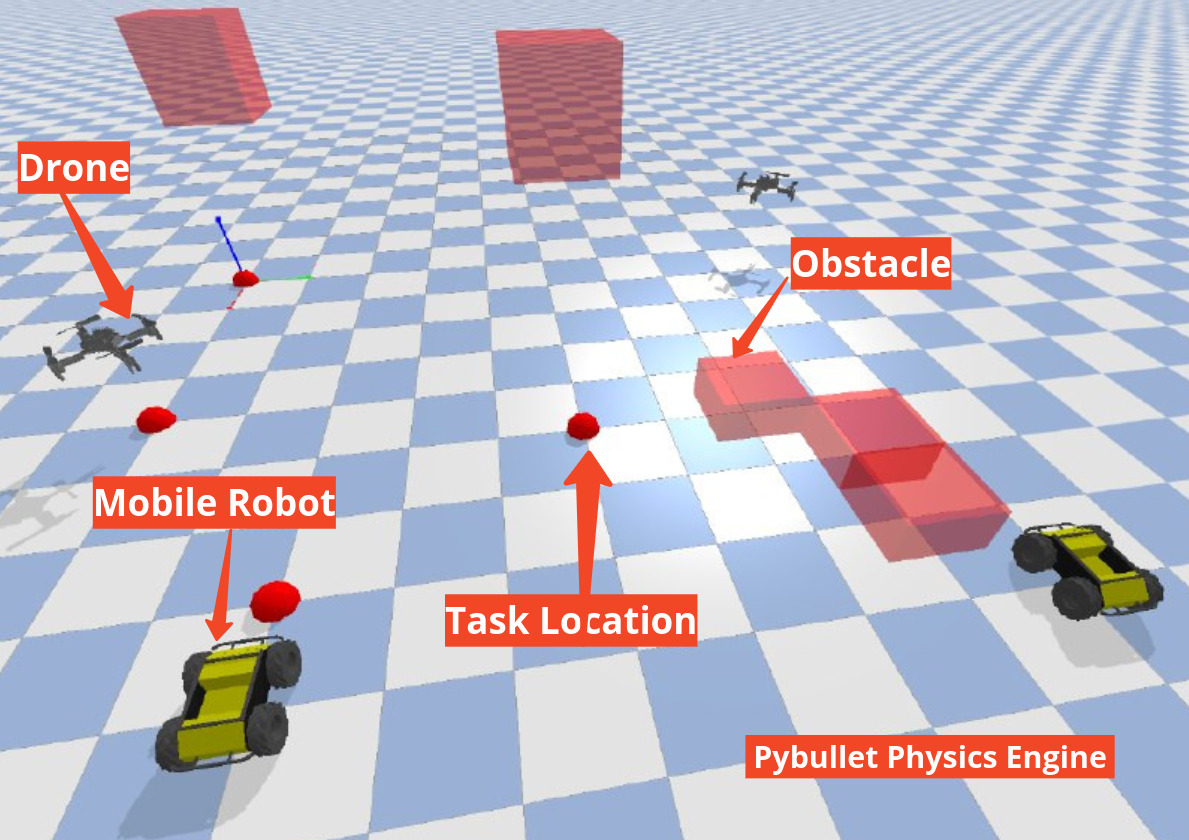}
 \caption{Simulation environment in PyBullet. Agents navigate with dynamic task assignments and obstacles.}
 \label{fig:sim_setup}
\end{figure}
Our custom simulator, built on PyBullet and integrated with Ray RLlib, models a dynamic three-dimensional grid-world in which mobile robots and drones navigate among obstacles and continuously generated tasks. 
Fig.~\ref{fig:sim_setup} shows our simulation environment. Random initialization of agent positions, task locations, and obstacles is performed at the start of each episode to ensure robustness.

Efficient navigation is achieved via an A* algorithm that computes the shortest-path distance \(d_{ij}(t)\) from an agent’s current position \(\mathbf{p}_i(t)\) to a task location \(\mathbf{l}_j\). The travel cost is computed as:
\begin{equation}
  c_{ij}(t) \;=\; \frac{d_{ij}(t)}{v_i},
\end{equation}
where $c_{ij}(t)$ is then subsequently clipped to the maximum value of the cost obtained and normalized to \([-1,1]\). To prevent collisions, a reservation-based conflict resolution mechanism is incorporated into the navigation process. Specifically, if an agent’s planned next cell is occupied either by an obstacle or by another robot, the agent does not proceed into that cell; instead, it triggers a re-planning process using the A* algorithm. If an agent experiences repeated blockages (exceeding a predefined threshold), its path is re-computed to adapt to the dynamic environment. This approach ensures that agents continually update their routes to avoid conflicts, thereby promoting efficient and collision-free task execution.

\section{Experiments}
\subsection{Experimental Setup}
\label{sec:experiments}
We conducted experiments in a $50 \times 50 \times 30$ m simulated environment using PyBullet, with a mix of ground robots (velocity $\sim 3$ m/s) and drones (velocity $\sim 5$ m/s). At any given time, the system handles 30 active tasks. As tasks are allocated, new tasks are added to the pool. The robots navigate to the tasks using the reservation-based A\(^*\) path planner.

We evaluated three key metrics: total travel time (Cost), which reflects allocation efficiency; conflict-free success rate, the fraction of tasks assigned to exactly one agent; and allocation time, i.e., the time taken for agents to reach an assignment decision.
We compare HIPPO-MAT with several baselines. The Hungarian Algorithm, a centralized optimal solution, serves as a reference to measure how closely HIPPO-MAT approximates optimal allocation. We also compare it with IPSO-G, a heuristic method, and Random Assignment, which shows how HIPPO-MAT improves travel time and conflict resolution. Additionally, HIPPO-MAT is compared to MAPPO+GCN, combining MAPPO with graph convolution networks (GCN), designed for fixed-task environments. Since HIPPO-MAT is for continuous task allocation, we evaluated its performance in dynamic environments where tasks are continuously generated and reassigned.

The scalability of the system was tested with up to 30 robots and continuously generated tasks, demonstrating computational efficiency through GraphSAGE. This approach provides efficient aggregation, making it more suitable for large-scale systems than attention-based models, which can be computationally expensive. GraphSAGE reduces complexity, making it ideal for real-time multi-agent task allocation in sparse graphs.

\subsection{Real-World Environment Setup}  
\label{sec:realworldvalidation}
\begin{figure}[ht]
 \centering
 \includegraphics[width=0.98\linewidth]{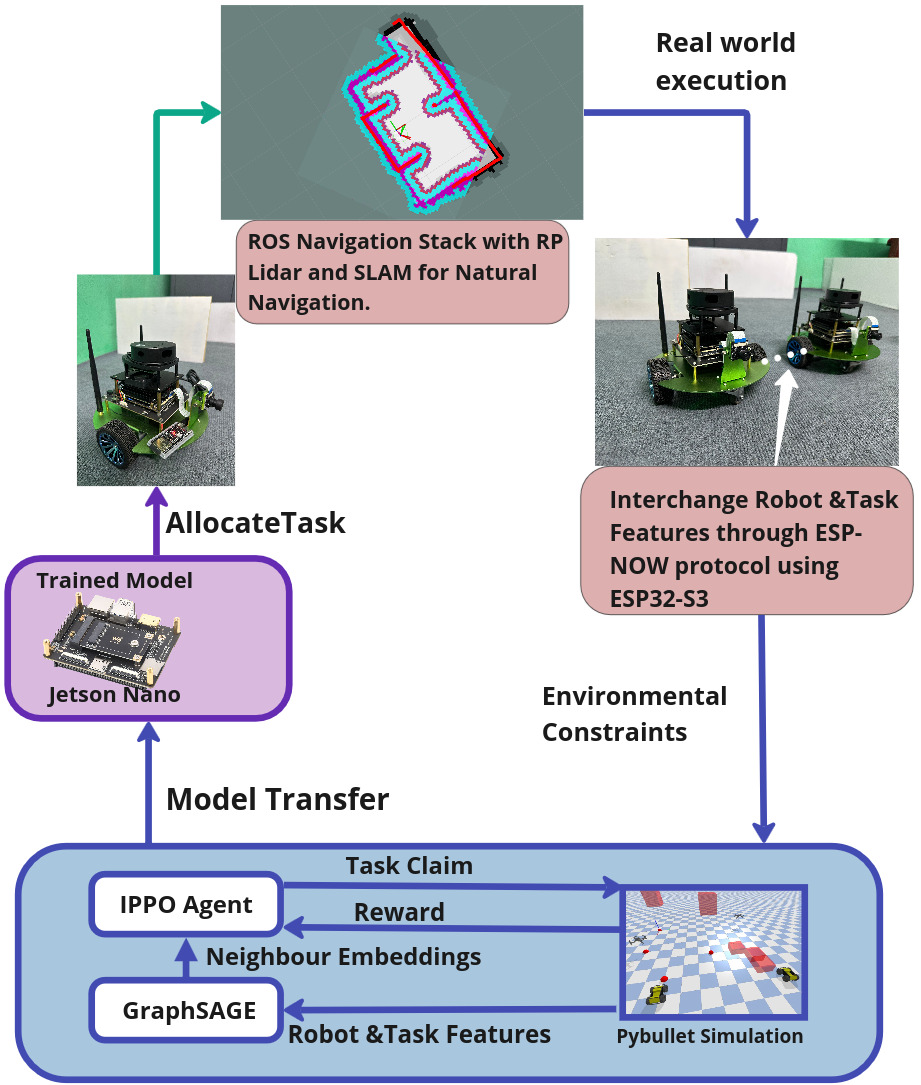}
 \caption{Experimental setup to test the model in Jetbot ROS AI robots equipped with Jetson Nano for individual policy loading and ESP32-S3 for inter-agent communication.}
 \label{fig:cover}
\end{figure}
\begin{figure}[ht]
 \centering
 \includegraphics[width=0.8\linewidth]{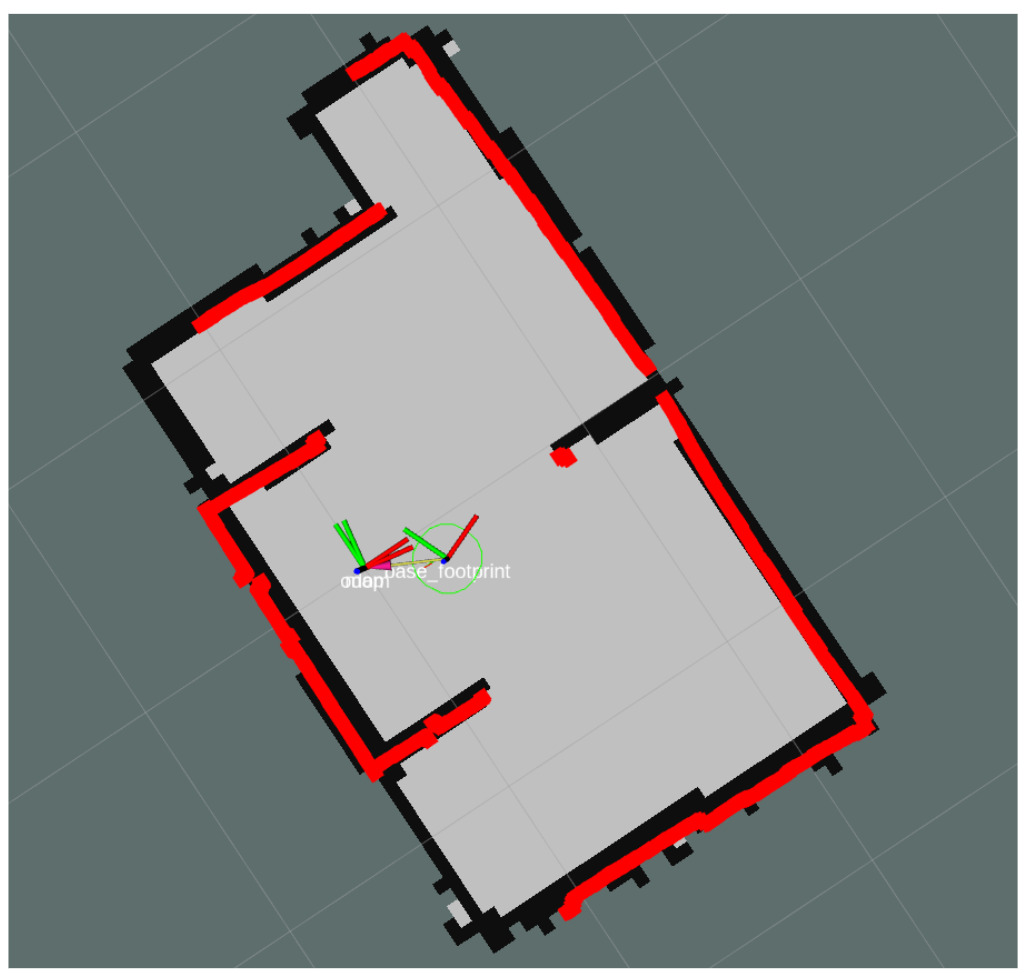}
 \caption{Environment mapping by Jetbot ROS AI robot.}
 \label{fig:mapping}
\end{figure}

To validate HIPPO-MAT, we performed real-world experiments using the JetBot ROS AI Robot, powered by Jetson Nano, with ESP32-S3 modules for communication. The setup shown in Fig.~\ref{fig:cover} enables the direct application of the task allocation model, trained in simulation, to real-world robots for real-time decision-making. The environment consists of JetBot robots assigned tasks generated at random locations. The robots navigate, complete tasks, and reallocate new tasks as they are generated. ESP32-S3 modules provide peer-to-peer communication using ESP-NOW protocol, where the Jetson Nano board was used to load an independent policy in each robot, allowing decentralized coordination without centralized control. Additionally, we utilize SLAM on-board of real robots to handle navigation. The environment mapped by the robots is shown in Fig.~\ref{fig:mapping}.

The real-world validation focused on testing how well HIPPO-MAT’s decentralized independent policy handles conflict resolution and generates optimal solutions. The experiments demonstrated that HIPPO-MAT efficiently resolves conflicts and allocates tasks, even with real-world sensor noise and communication delays. GraphSAGE ensures computational efficiency and robust performance in dynamic task allocation scenarios, confirming HIPPO-MAT’s reliability for continuous task allocation in decentralized multi-agent systems.

\subsection{Results and Discussion}
\subsubsection{Training performance}
\begin{figure}[ht]
 \centering
 \smallskip
 \includegraphics[width=0.45\textwidth]{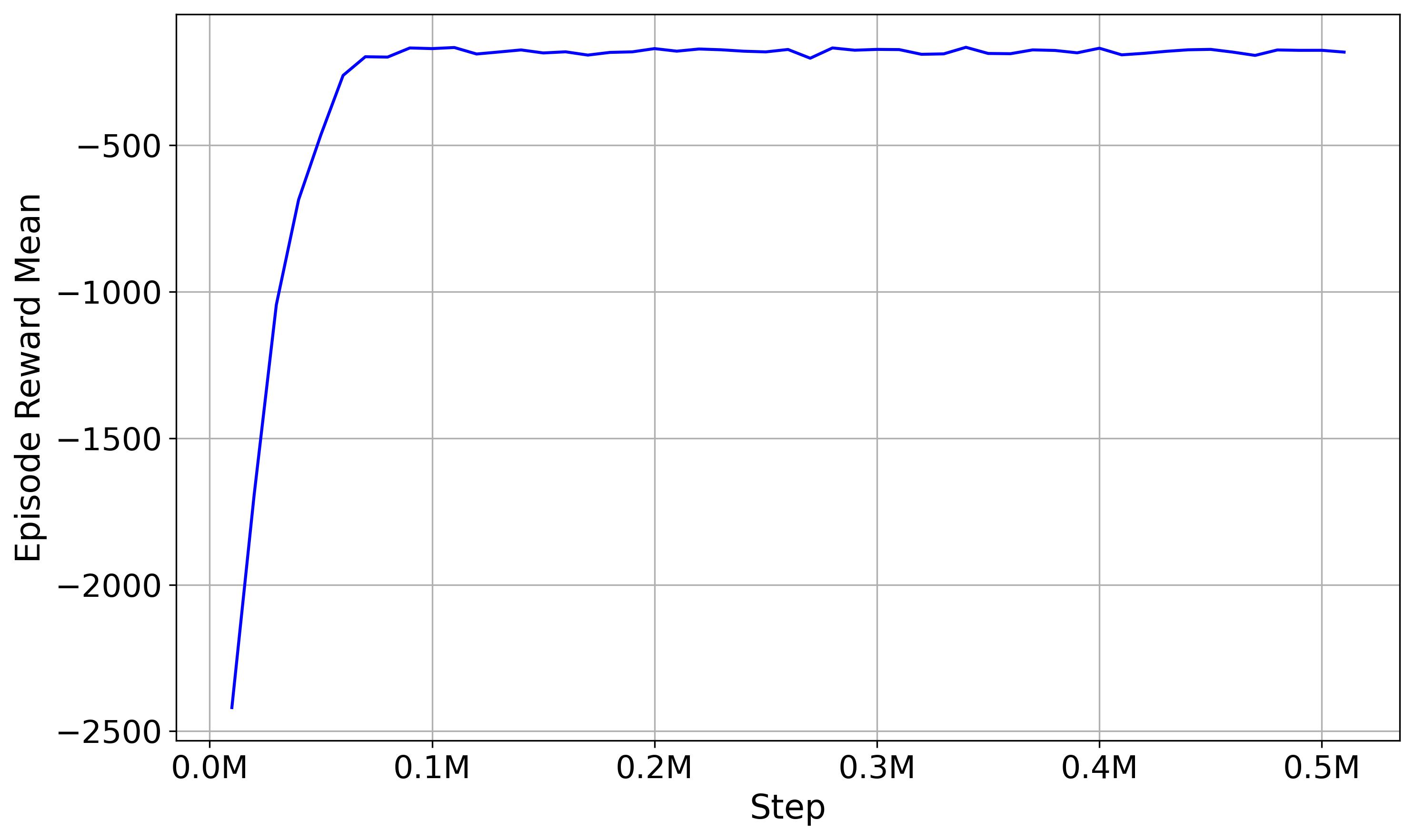}
 \caption{Mean reward vs. training steps during learning.}
 \label{fig:reward}
\end{figure}

\begin{figure}[tp]
 \centering
 \smallskip
 \includegraphics[width=0.45\textwidth]{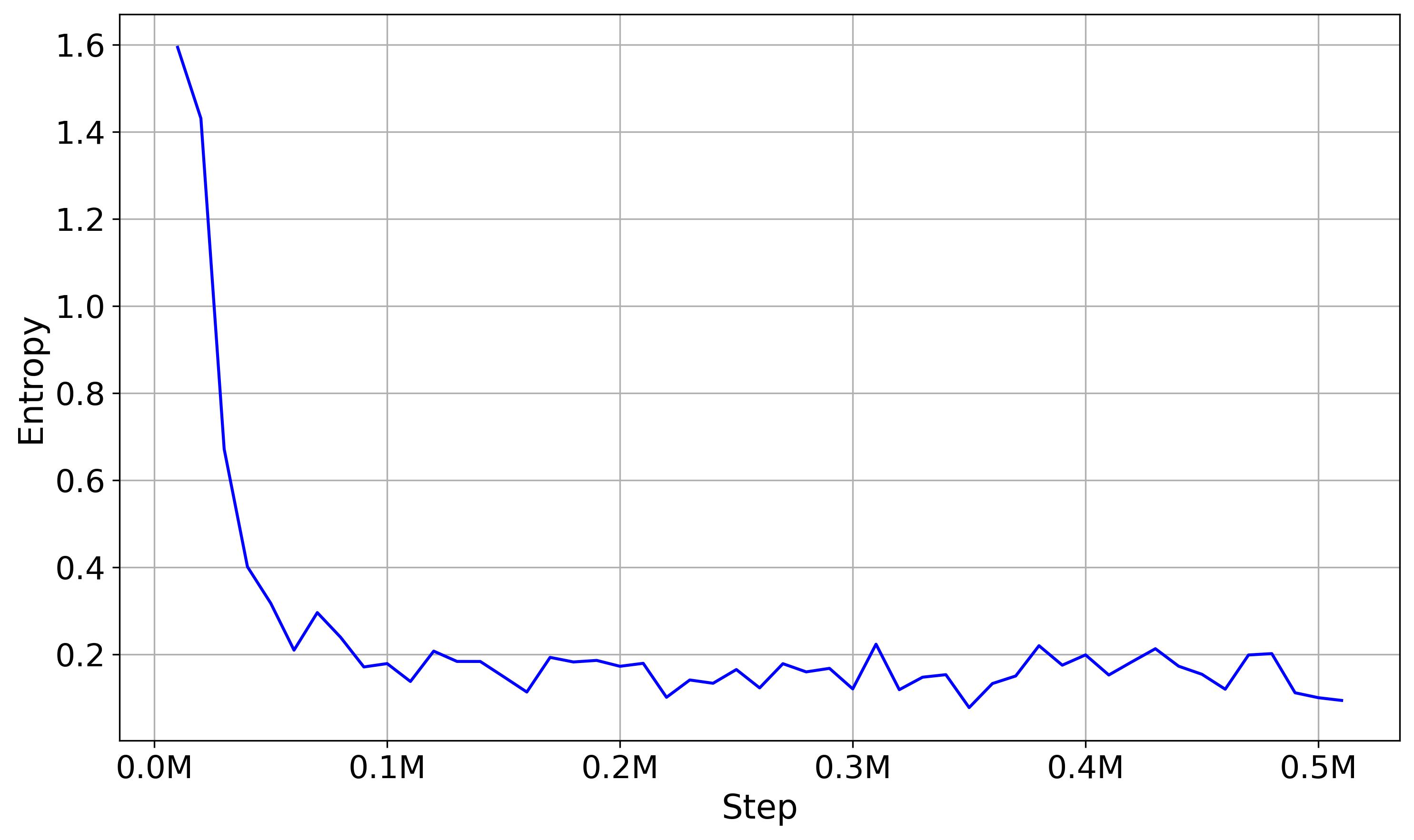}
 \caption{Entropy vs. training steps during learning.}
 \label{fig:entropy}
\end{figure}
The increase in rewards shown in Fig.~\ref{fig:reward}, coupled with a decrease in entropy shown in Fig.~\ref{fig:entropy}, indicates that agents are progressively learning to allocate tasks more efficiently over time. Initially, high entropy reflects uncertainty in decision-making due to insufficient experience, leading to suboptimal task assignments and lower rewards. However, as training progresses, agents refine their policies, resulting in more consistent and confident task selection. This is reflected in the rising reward trend, which signifies that agents are successfully maximizing their task allocation efficiency. The entropy reduction further confirms that agents are transitioning from an exploratory phase to a more deterministic and optimized decision-making process, ultimately leading to improved overall system performance.

\subsubsection{Cost of Total Travel Time}
We evaluate the total travel cost in both fixed-task and continuous-task allocation scenarios. In the fixed-task scenario, tasks are static not continuous, we only generated at one time. In the continuous-task scenario, tasks are dynamically replenished as robots allocating existing ones. The total travel costs for both scenarios are summarized in Table~\ref{tab:comp_fixed} (for fixed tasks) and Table~\ref{tab:comp_continuous} (for continuous tasks).


\begin{table}[!ht]
\centering
\caption{Total Travel Cost in Fixed Task Allocation vs. Number of Agents.}
\label{tab:comp_fixed}
\begin{tabular}{|l|*{4}{c|}}
\hline
\backslashbox{\textbf{Method}}{\textbf{N Agents}} 
 & \makebox[2em]{\textbf{4}}
 & \makebox[2em]{\textbf{8}}
 & \makebox[2em]{\textbf{12}}
 & \makebox[2em]{\textbf{20}} \\
\hline
Hungarian  & 20.3 & 60.3 & 131.9 & 254.3 \\
HIPPO-MAT  & 20.3 & 62.1 & 138.2 & 327.4 \\
MAPPO+GCN  & 20.3 & 61.2 & 134.7 & 321.5 \\
IPSO-G     & 22.5 & 65.8 & 140.5 & 383.2 \\
Random     & 27.9 & 72.3 & 175.7 & 423.8 \\
\hline
\end{tabular}
\end{table}

In the fixed-task scenario, we evaluated the performance with varying number of robots count; HIPPO-MAT consistently outperforms IPSO-G and Random Assignment, with performance approaching that of the Hungarian algorithm MAPPO+GCN, while effective for fixed tasks.


 
\begin{table}[!ht]
\smallskip
\centering
\caption{Total Travel Cost in Continuous Task Allocation vs. Number of Agents.}
\label{tab:comp_continuous}
\begin{tabular}{|l|*{4}{c|}}
\hline
\backslashbox{\textbf{Method}}{\textbf{N Agents}} 
 & \makebox[2em]{\textbf{5}}
 & \makebox[2em]{\textbf{10}}
 & \makebox[2em]{\textbf{20}}
 & \makebox[2em]{\textbf{30}} \\
\hline
Hungarian & 281.58 & 560.3 & 1931.16 & 4524.3 \\
HIPPO-MAT & 295.58 & 612.1 & 2236.2 & 6212.4 \\
MAPPO+GCN & 295.3 & 638.4 & 2342.0 & 6701.2 \\
IPSO-G & 302.3 & 667.2 & 2445.0 & 7201.1 \\
Random & 484.4 & 736.5 & 3282.3 & 9453.2 \\
\hline
\end{tabular}
\end{table}

For continuous-task allocation, HIPPO-MAT continues to outperform IPSO-G and Random Assignment, showing robust performance even as tasks are dynamically added. MAPPO+GCN, although effective for fixed tasks, struggles with the added complexity of continuously replenished tasks, resulting in conflicting solutions, and compared to our method, it shows a lower performance.

\subsubsection{Conflict-Free Success Rate and Allocation Time}
We measured the conflict-free success rate, defined as the percentage of tasks successfully assigned to exactly one agent. In decentralized task allocation, where all robots access tasks simultaneously, handling conflicts is the primary challenge, most of the studies do not consider this metric. Additionally, we measured the total time for a decision to completely allocate the fixed number of tasks. Tables~\ref{tab:success_rate_fixed} and~\ref{tab:success_rate_continuous} show the conflict-free success rate for fixed and continuous task allocation scenarios, respectively.

\begin{table}[!ht]
\centering
\caption{Conflict-Free Success Rate and Allocation Time for Fixed Task Allocation vs. Number of Agents.}
\label{tab:success_rate_fixed}
\begin{tabular}{|c|c|c|c|c|c|}
\hline
\textbf{Metric} & \textbf{Method} & \textbf{5} & \textbf{10} & \textbf{20} & \textbf{30} \\
\hline
\multirow{3}{*}{\textbf{Success Rate (\%)}} & Hungarian & 100 & 100 & 100 & 100 \\
& HIPPO-MAT & 100 & 100 & 90 & 80 \\
& MAPPO+GCN & 100 & 90 & 70 & 70 \\
& IPSO-G & 90 & 80 & 80 & 60 \\
\hline
\multirow{3}{*}{\textbf{Allocation Time (s)}} & Hungarian & 0.8 & 1.5 & 2.8 & 5.6 \\
& HIPPO-MAT & 0.2 & 0.2 & 0.4 & 0.5 \\
& MAPPO+GCN & 0.4 & 0.6 & 1.5 & 3.2 \\
& IPSO-G & 0.3 & 0.3 & 0.5 & 1.2 \\
\hline
\end{tabular}
\end{table}

\begin{table}[!ht]
\centering
\caption{Conflict-Free Success Rate for Continuous Task Allocation  vs. Number of Agents.}
\label{tab:success_rate_continuous}
\begin{tabular}{|c|c|c|c|c|c|}
\hline
\textbf{Metric} & \textbf{Method} & \textbf{5} & \textbf{10} & \textbf{20} & \textbf{30} \\
\hline
\multirow{3}{*}{\textbf{Success Rate (\%)}} & Hungarian & 100 & 100 & 100 & 100 \\
& HIPPO-MAT & 100 & 100 & 90 & 80 \\
& MAPPO+GCN & 60 & 60 & 50 & 50 \\
& IPSO-G & 80 & 80 & 70 & 60 \\
\hline
\end{tabular}
\end{table}

As shown in the tables for both scenarios, HIPPO-MAT maintains a high success rate for conflict-free task allocation, outperforming IPSO-G and Random Assignment. While Hungarian achieves a perfect success rate. But the CTDE approach, MAPPO+GCN achieves lower success rates as the number of dynamically generated tasks increases. Compare to centralized hungarian and other approach, our HIPPO-MAT perform faster to make decision.

\subsubsection{Ablation Study: No GraphSAGE (Only Local Information)}
We also performed an ablation study to evaluate the performance of HIPPO-MAT without GraphSAGE. In this study, the model only used local agent observations without aggregating information from neighboring agents. The results showed that while the model could still allocate tasks, it performed significantly worse in terms of scalability and conflict resolution compared to the full HIPPO-MAT model with GraphSAGE. This highlights the importance of GraphSAGE for information aggregation and the ability to scale efficiently in large multi-agent systems.

\section{Conclusion and Future Work}
\label{sec:conclusion}
In this paper, we introduced HIPPO-MAT, a novel decentralized task allocation framework for multi-agent systems that integrates GraphSAGE-based graph neural networks with an independent PPO (IPPO) algorithm. Our approach is tailored for continuous task allocation in dynamic 3D environments, enabling heterogeneous agents—such as unmanned aerial vehicles (UAVs) and unmanned ground vehicles (UGVs)—to concurrently access and assign tasks without relying on centralized coordination. Our method shows GraphSAGE to generate enriched state embeddings for each agent by aggregating information from all other agents. These embeddings are then used by an independent policy network trained via IPPO to make rapid, conflict-aware task assignment decisions. A modified reservation-based A* path planner shows efficient routing and collision avoidance during task execution.

Experimental evaluations demonstrate that HIPPO-MAT achieves near-optimal performance relative to the centralized Hungarian algorithm. In the fixed task allocation scenario, the travel cost gap averages approximately 9.1\% higher than that of the centralized approach, while in the continuous task allocation scenario the gap increases to about 16.9\%. Despite these modest increases in travel cost, our framework offers a dramatic reduction in allocation time—up to 90\% faster—and maintains a high conflict-free success rate of 92.5\%, outperforming baseline methods such as IPSO-G and MAPPO+GCN. Real-world experiments on JetBot ROS AI Robots further validate the practicality of HIPPO-MAT. These tests confirm that our decentralized framework can robustly handle sensor noise and communication delays, efficiently resolving conflicts and consistently allocating tasks in real time using Jetson Nano and ESP32-S3 with ESP-NOW communication protocol.

Future work will focus on refining the reward structure and integrating attention mechanisms to further improve conflict resolution in highly dynamic environments. Additionally, we aim to extend our framework to larger-scale systems and investigate its integration with advanced SLAM techniques to enhance navigation and mapping capabilities in complex operational scenarios. 

\bibliography{references}

\begin{thebibliography}{99}

\bibitem{Kuhn_1955}H. W. Kuhn, “The hungarian method for the assignment problem,” Naval Research Logistics Quarterly, vol. 2, no. 1-2, pp. 83–97, 1955.

\bibitem{graphsage}
W. L. Hamilton, R. Ying, and J. Leskovec, “Inductive representation learning on large graphs," in \textit{Proceedings of the 31st International Conference on Neural Information Processing Systems}, Long Beach, California, USA, 2017, pp. 1025-1035.

\bibitem{goarin2024graph} M. Goarin and G. Loianno, “Graph Neural Network for Decentralized Multi-Robot Goal Assignment," in \textit{IEEE Robotics and Automation Letters}, vol. 9, no. 5, pp. 4051-4058, May 2024.

\bibitem{ismail2017decentralized}
S. Ismail and L. Sun, “Decentralized Hungarian-based approach for fast and scalable task allocation," in \textit{2017 International Conference on Unmanned Aircraft Systems (ICUAS)}, Miami, FL, USA, 2017, pp. 23-28.

\bibitem{xia}
M. Xia, ``Research on decentralized task allocation and collaboration based on multiple AUVs,'' in \textit{2023 3rd International Signal Processing, Communications and Engineering Management Conference (ISPCEM)}, Montreal, QC, Canada, 2023, pp. 274-281.

\bibitem{greedy}
X. Kong, Y. Gao, T. Wang, J. Liu, and W. Xu, ``Multi-robot task allocation strategy based on particle swarm optimization and greedy algorithm,'' in \textit{2019 IEEE 8th Joint International Information Technology and Artificial Intelligence Conference (ITAIC)}, Chongqing, China, 2019, pp. 1643-1646.

\bibitem{genetic}
M. Shelkamy, C. M. Elias, D. M. Mahfouz, and O. M. Shehata, ``Comparative analysis of various optimization techniques for solving multi-robot task allocation problem,'' in \textit{2020 2nd Novel Intelligent and Leading Emerging Sciences Conference (NILES)}, Giza, Egypt, 2020, pp. 538-543.

\bibitem{neurofleets}
R. Peter, L. Ratnabala, E. Y. Andrew Charles, and D. Tsetserukou, ``Dynamic subgoal based path formation and task allocation: A NeuroFleets approach to scalable swarm robotics,'' in \textit{2024 IEEE International Conference on Systems, Man, and Cybernetics (SMC)}, Kuching, Malaysia, 2024, pp. 878-883.

\bibitem{zhong2018stable}
Y. Zhong, L. Gao, T. Wang, S. Gong, B. Zou, and D. Yu, ``Achieving stable and optimal passenger-driver matching in ride-sharing system,'' in \textit{2018 IEEE 15th International Conference on Mobile Ad Hoc and Sensor Systems (MASS)}, Chengdu, China, 2018, pp. 125-133.

\bibitem{liu2024multi}
J. Liu, Z. Zhang, T. Xu, and C. Yan, ``Multi-UAV united task allocation via extended market mechanism based on flight path cost,'' in \textit{2024 IEEE International Conference on Unmanned Systems (ICUS)}, Nanjing, China, 2024, pp. 50-55.

\bibitem{harmonyDTA}
M. Eser and A. E. Yilmaz, ``A gossip-based auction algorithm for decentralized task rescheduling in heterogeneous drone swarms,'' in \textit{IEEE Transactions on Aerospace and Electronic Systems}.

\bibitem{agrawal2023rtaw}
A. Agrawal, A. S. Bedi, and D. Manocha, ``RTAW: An attention inspired reinforcement learning method for multi-robot task allocation in warehouse environments,'' in \textit{2023 IEEE International Conference on Robotics and Automation (ICRA)}, London, United Kingdom, 2023, pp. 1393-1399.

\bibitem{dai2025scheduling}
W. Dai, U. Rai, J. Chiun, Y. Cao, and G. Sartoretti, ``Heterogeneous multi-robot task allocation and scheduling via reinforcement learning,'' \textit{IEEE Robotics and Automation Letters}, vol. 10, no. 3, pp. 2654-2661, 2025.

\bibitem{mappo2023arxiv} Z.Da, ``Research on multi-agent communication and collaborative decision-making based on deep reinforcement learning," 2023, arXiv:2305.17141v1.

\bibitem{lowe2017multi}
R. Lowe, Y. Wu, A. Tamar, J. Harb, O. P. Abbeel, and I. Mordatch, ``Multi-agent actor-critic for mixed cooperative-competitive environments,'' in \textit{Advances in Neural Information Processing Systems 30 (NeurIPS 2017)}, Long Beach, CA, USA, 2017, pp. 6382--6393.

\bibitem{ippo}
C. Schr{\"o}der de Witt, T. Gupta, D. Makoviichuk, V. Makoviychuk, P. H. S. Torr, M. Sun, and S. Whiteson, ``Is independent learning all you need in the StarCraft multi-agent challenge?,'' \textit{CoRR}, vol. abs/2011.09533, 2020.

\bibitem{peng2024graph}
J. Peng, H. Viswanath, and A. Bera, ``Graph-based decentralized task allocation for multi-robot target localization,'' \textit{IEEE Robotics and Automation Letters}, vol. 9, no. 11, pp. 10676-10683, 2024.

\bibitem{blumenkamp2022framework}
J. Blumenkamp, S. Morad, J. Gielis, Q. Li, and A. Prorok, ``A framework for real-world multi-robot systems running decentralized GNN-based policies,'' in \textit{Proc. 2022 International Conference on Robotics and Automation (ICRA)}, 2022, pp. 8772-8778.

\bibitem{Hetgppo}
M. Bettini, A. Shankar, and A. Prorok, ``Heterogeneous multi-robot reinforcement learning,'' in \textit{Proc. 2023 International Conference on Autonomous Agents and Multiagent Systems (AAMAS)}, London, United Kingdom, 2023, pp. 1485-1494.

\bibitem{Ratnabala_2025}
L. Ratnabala, A. Fedoseev, R. Peter, and D. Tsetserukou, ``MAGNNET: Multi-agent graph neural network-based efficient task allocation for autonomous vehicles with deep reinforcement learning,'' 2025, arXiv:2502.02311.

\balance
\end{thebibliography}

\end{document}